\documentclass[runningheads]{cl2emult}

\usepackage{makeidx}  
\usepackage{graphicx} 
\usepackage{subeqnar} 
\usepackage{multicol} 
\usepackage{cropmark} 
\usepackage{eso}      
\makeindex            

\begin{document}
\title*{Evolution of early-type galaxies in clusters}
\toctitle{Evolution of early-type galaxies in clusters}
%
%
\titlerunning{Early-type galaxies in clusters}
%
\author{Pieter van Dokkum\inst{1}
\and Marijn Franx\inst{2}
\and Daniel Kelson\inst{3}
\and Garth Illingworth\inst{4}
\and Daniel Fabricant\inst{5}}
\authorrunning{Pieter van Dokkum et al.}
%
%
\institute{California Institute of Technology, Pasadena CA 91125, USA
\and Leiden Observatory, PO Box 9513, 2300 RA Leiden, Netherlands
\and OCIW, Pasadena CA 91101, USA
\and UCO/Lick Observatory, Santa Cruz CA 95064, USA
\and Center for Astrophysics, Cambridge MA 02318, USA}

\maketitle              

\begin{abstract}

The slow evolution of the $M/L$ ratios, colors, and line strengths of
cluster early-type galaxies to $z\sim 1$ suggests that their stars
were formed at very high redshift.  At the same time, morphological
studies of distant clusters indicate significant evolution in the
early-type galaxy population.  Striking evidence for strong
morphological evolution in clusters is the discovery of a large number
of red merger systems in the cluster MS\,1054--03 at $z=0.83$.  The
presence of these mergers is qualitatively consistent with predictions
from hierarchical galaxy formation models, and is direct evidence
against an early collapse for all early-type galaxies.  In most of the
mergers there is no evidence for strong star formation.  Therefore the
mean stellar ages of the merger products will be much older than the
``assembly age'', and do not violate the constraints on the star
formation epoch of early-type galaxies imposed by the color-magnitude
relation and the Fundamental Plane.

\end{abstract}

\section{Introduction}
The field of galaxy evolution in rich clusters has witnessed great
progress in recent years, in large part because of the powerful
combination of Hubble Space Telescope imaging and spectroscopy with
large ground-based telescopes. Examples of succesful programs are the
MORPHS collaboration \cite{smail}, who obtained deep HST images of
$\sim 10$ clusters at $0.3<z<0.5$, and the CNOC group \cite{yee}, who
obtained extensive wilde field spectroscopy and imaging of X-ray
selected clusters at $0.2<z<0.5$.

Our strategy is complementary to these efforts. We are obtaining wide
field, deep HST WFPC2 images of several clusters at $0.3<z<0.9$, in
combination with extensive spectroscopic follow-up using the Keck
telescope.  The combination of the various programs will produce a
balanced data set on distant galaxy clusters.

\section{Evidence for old stellar populations}
Studies of the evolution of early-type galaxies in clusters are
in remarkable agreement. Studies of colors \cite{ellis,stanford},
$M/L$ ratios \cite{vD96,K97,vD98,bender,K00}, and
line strenghts \cite{bender,K01}
of early-type galaxies in distant clusters all
show that they form a very homogeneous, slowly evolving
population all the way from $z=0$ to $z=1$. As an example, the scatter
in the color-magnitude relation remains very small
to $z \sim 1$ \cite{stanford}, placing a strong upper limit on the
scatter in ages of early-type galaxy at any given time.

The strongest constraints on the {\em mean} star formation epoch have
come from the evolution of the
Fundamental Plane (FP) relation.
The Fundamental Plane \cite{george}
is a relation between
the effective radius $r_e$, effective surface brightness $\mu_e$, and
central velocity dispersion $\sigma$, such that $ r_e \mu_e^{0.8}
\propto \sigma^{1.25}$.
The implication of the existence of the Fundamental Plane is
that $M/L$ ratios of galaxies correlate strongly with their structural
parameters: $M/L \propto r_e^{0.2}\sigma ^{0.4} \propto M^{0.2}$
\cite{faber}.
The power of the FP lies in its small scatter and its relation
to $M/L$ ratios.
The $M/L$ ratios of galaxies evolve because the luminosity of
their stellar populations decreases as they age (``passive
evolution''), and the evolution of the intercept of the
FP gives a strong constraint on the mean
stellar age of early-type galaxies.

\begin{figure}
\centering
\includegraphics[width=1.0\textwidth]{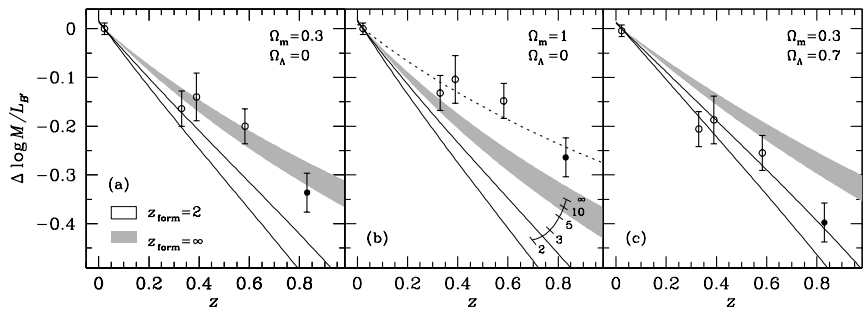}
\caption[]{Evolution of the mean $M/L_B$ ratio of early-type
galaxies, as determined from the Fundamental Plane relation
\cite{vD98}.
The observed evolution is slow, indicating a high formation
redshift of the stars.}
\label{fp.plot}
\end{figure}

The measured evolution of $M/L_B$ to $z=0.83$ is shown in Fig.\ 
\ref{fp.plot}, from \cite{vD98}.
The evolution is surprisingly low, $\ln M/L_B \propto -z$,
indicating stellar formation redshifts of $z > 3$ for $\Omega_m=0.3$
and $\Omega_{\Lambda}=0$ \cite{vD98}.

\section{Evidence for morphological evolution}

Although early-type galaxies appear to form a very stable, slowly evolving
population, evidence is accumulating that a large fraction
of early-type
galaxies in nearby clusters was relatively recently assembled,
and/or transformed from spiral galaxies. Hence the (usually implicit)
assumption in studies of the CM relation or the FP
that the set of high redshift early-type
galaxies is similar to the
set of low redshift early-type galaxies is probably not justified.

\subsection{Evolution of the early-type galaxy fraction}

\begin{figure}
\centering
\includegraphics[width=0.4\textwidth]{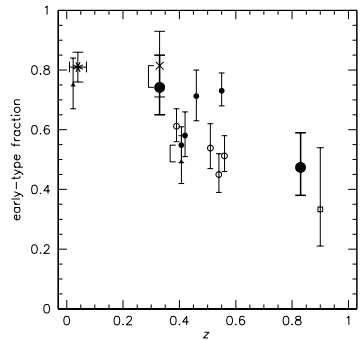}
\caption[]{Evolution of the early-type galaxy fraction in clusters
\cite{vD00}. Except CL\,1358+62 at $z=0.33$
and MS\,1054--03 at $z=0.83$ data are from \cite{dressler}
and \cite{lubin}.}
\label{early.plot}
\end{figure}

Dressler et al.\ report a high fraction of spiral
galaxies in clusters at $0.3< z < 0.5$ \cite{dressler}. These galaxies
are nearly absent in nearby rich clusters, and hence
must have transformed into early-type galaxies between $z=0.5$ and
$z=0$. Other studies \cite{couch,vD00}
have confirmed this trend, and extended it to $z=0.8$.
The evolution of the early-type galaxy fraction
is shown in Fig.\ \ref{early.plot}. The early-type fraction
decreases by a factor $\sim 2$  from $z=0$ to $z=1$, although
the trend has significant scatter.

Dressler et al.\ found that the increased fraction of spiral galaxies
at high redshift is accompanied by a low fraction of S0 galaxies, and
concluded that the $z \approx 0.4$ spiral galaxies transform into S0
galaxies \cite{dressler}.  Other studies have found evidence for
merging and interactions in high redshift clusters
\cite{lpm,dressler94,couch,vD99}. These transformations may provide a
way to form young elliptical galaxies in the clusters at late times.

\subsection{Mergers in MS\,1054--03 at $z=0.83$}

The discovery of a large number of red merger systems in the cluster
MS\,1054--03 at $z=0.83$ is arguably the most spectacular evidence for
recent formation of massive early-type galaxies.  We have obtained
deep, multi-color images of MS\,1054--03 at 6 pointings with WFPC2 on
HST. The Keck telescope was used to measure redshifts of 186 galaxies,
and 80 of those are cluster members.  Together with data from the
literature, we found 89 cluster members, 81 of which are
in the HST mosaic.
We classified the galaxies along the revised Hubble sequence,
allowing for a separate category of mergers. We combined classifications
of three of us, and verified that the results were robust from
classifier to classifier. The results have been presented in
\cite{vD99,vD00}.

\begin{figure}
\centering
\includegraphics[width=0.92\textwidth]{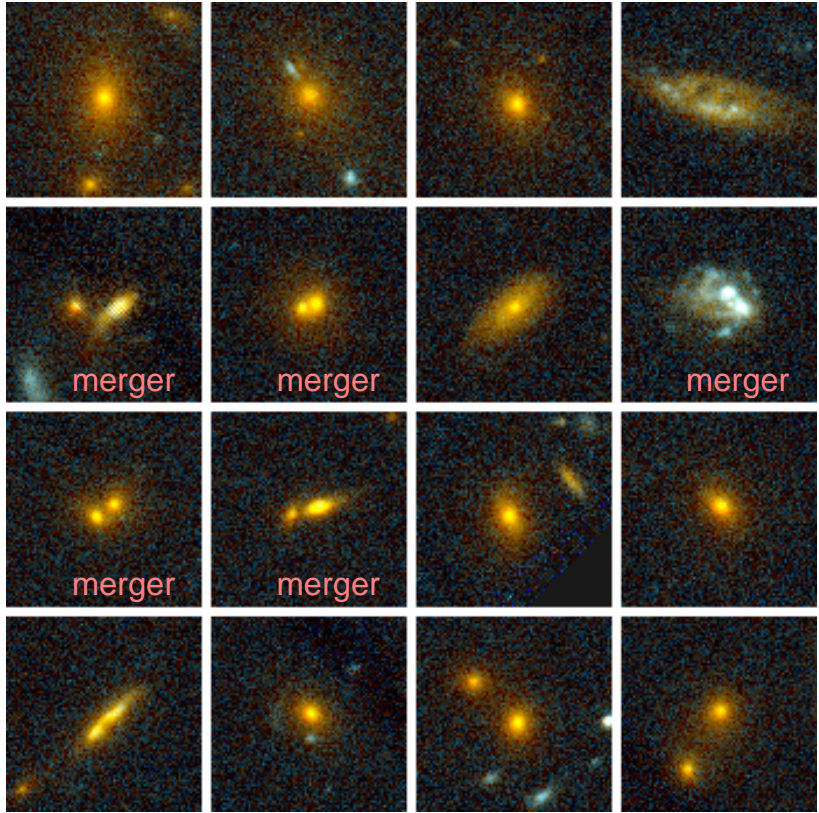}
\caption[]{The sixteen brightest confirmed members
of MS\,1054--03 at $z=0.83$, ordered by $I_{F814W}$ magnitude.
Note the large number of mergers.
A color version of this figure can
be found at http://www.astro.caltech.edu/\~{}pgd/ms1054.
}
\label{panel.plot}
\end{figure}

The most surprising result of our survey of MS\,1054--03
is the high fraction of mergers. Most of the mergers are 
very luminous ($M_B \sim -22$ in the rest frame,
or $\sim 2L_*$ at $z=0.83$), and a striking way to
display our result is to show a panel with the 16 brightest confirmed
cluster members (Fig.\ \ref{panel.plot}). Five out of these 16
were classified as mergers.

The mergers are generally red, with a few
exceptions. Similarly, the spectra of most of the mergers do not show
strong emission lines.
These results suggest that the bulk of the stars of the mergers were
formed well before the merger. Hence the stellar age of the merged
galaxies will be significantly different from the ``assembly age'', i.e.,
the time that has elapsed since the galaxy ``was put together''.

The colors of the mergers
are consistent with the hypothesis that they will evolve
into early-type galaxies.  After aging of the stellar
populations, the scatter of the total population of
mergers\,+\,early-type galaxies
will be very similar to the measured scatter in the CM relation
at $z<0.6$ \cite{vD00}.
Hence a low scatter at $z=0$ does not mean that all
galaxies in the population are homogeneous and very old: the influence
of merging can be small if the star formation involved with the
merging is low.

The physical reason for the low star formation is unknown: it is
possible that the massive precursor galaxies had already lost their
cold gas due to internal processes (such as super winds, or winds
driven by nuclear activity). Alternatively, the cluster environment
may play an important role: the cold gas may have been stripped by the
cluster X-ray gas.  Observations of more clusters may shed further
light on this effect.

\subsection{Close pairs}

\begin{figure}
\centering
\includegraphics[width=0.45\textwidth]{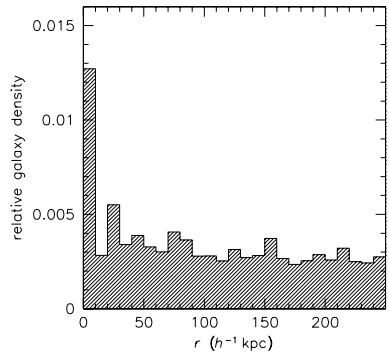}
\caption[]{
The overdensity of pairs in the outskirts of MS\,1054--03.  There is a
clear excess of pairs at small separations $< 10 h^{-1}$ kpc.  This is
independent confirmation of the enhanced interaction rate in
MS\,1054--03.}
\label{pairs.plot}
\end{figure}

Visual classifications are subjective in their nature.
An objective way to 
establish whether interactions and mergers were more prevalent in
high redshift clusters is to investigate the distribution of
galaxy -- galaxy separations. Figure \ref{pairs.plot} shows the
average galaxy density around red galaxies in our HST mosaic of
MS\,1054--03, excluding the central $200h^{-1}$ kpc.
The number of galaxies in each bin is weighted by
the area; therefore, a flat distribution would show that the
galaxies in the outer parts of the cluster are distributed uniformly.

There is clearly an excess of pairs at small separations ($<10
h^{-1}$\,kpc).  This result confirms the large number of interacting
galaxies in this cluster, and is completely independent from the
morphological classifications.  Only $\sim 50$\,\% of the close pairs
were classified as mergers. The other half may constitute a reservoir
of ``future'' mergers.  It will be interesting to measure the velocity
differences of paired galaxies, to confirm their physical association.

\section{Discussion}

The presence of the mergers is direct evidence against an early
collapse of all massive early-type galaxies at very high redshift,
and is in qualitative agreement with hierarchical galaxy formation
models. The main uncertainty is the assumption that the galaxy
population in MS\,1054--03 is typical for its redshift. It is
important to test whether other clusters at similar
redshift also show enhanced interaction rates in their outskirts.
It may be that such a phase of enhanced merging occurs at
different redshifts for different clusters. In MS\,1054--03 the
mergers probably occur in infalling subclumps,
and its high merger fraction could be related to its unvirialized
state \cite{vD99}.

It may seem difficult to reconcile the homogeneity and slow evolution
of early-type galaxies with the strong evolution
inferred from morphological studies. However, the observations
of MS\,1054--03 clearly show that the sample of early-type galaxies
at high redshift is only a subset of the sample of low redshift
early-type galaxies. Studies of the evolution of
early-type galaxies systematically discard the youngest progenitors
of present-day early-types, and this leads to a biased estimate
of their age. We have dubbed this effect the ``progenitor bias''
\cite{FvD96,vD96}, and we model its effects on the observed
evolution of early-type galaxies in detail
in \cite{vDF00}.

\clearpage
\addcontentsline{toc}{section}{Index}
\flushbottom
\printindex

\end{document}